\documentclass[10pt]{article} 
\usepackage[preprint]{tmlr}

\usepackage[colorlinks=true]{hyperref}
\usepackage{url}

\usepackage{tabularray}
\UseTblrLibrary{siunitx}

\usepackage{booktabs}
\usepackage{multirow}
\usepackage{multicol}
\usepackage{enumitem}
\usepackage{rotating}
\usepackage{algorithm}
\usepackage{colortbl}
\usepackage{xcolor}
\usepackage{subfig}

\usepackage{tepper}

\usepackage[style=numeric,sortcites=true]{biblatex}
\bibliography{refs}

\title{GleanVec: Accelerating vector search with minimalist nonlinear dimensionality reduction}

\author{%
    \name Mariano Tepper \email mariano.tepper@intel.com \\
    \name Ishwar Singh Bhati \email ishwar.s.bhati@intel.com \\
    \name Cecilia Aguerrebere \email cecilia.aguerrebere@intel.com\\
    \name Ted Willke \email ted.willke@intel.com\\
    \addr Intel Labs\\
}

\begin{document}

\maketitle

\begin{abstract}
Embedding models can generate high-dimensional vectors whose similarity reflects semantic affinities. Thus, accurately and timely retrieving those vectors in a large collection that are similar to a given query has become a critical component of a wide range of applications. In particular, cross-modal retrieval (e.g., where a text query is used to find images) is gaining momentum rapidly. Here, it is challenging to achieve high accuracy as the queries often have different statistical distributions than the database vectors. Moreover, the high vector dimensionality puts these search systems under compute and memory pressure, leading to subpar performance.
In this work, we present new linear and nonlinear methods for dimensionality reduction to accelerate high-dimensional vector search while maintaining accuracy in settings with in-distribution (ID) and out-of-distribution (OOD) queries. The linear LeanVec-Sphering outperforms other linear methods, trains faster, comes with no hyperparameters, and allows to set the target dimensionality more flexibly. The nonlinear Generalized LeanVec (GleanVec) uses a piecewise linear scheme to further improve the search accuracy while remaining computationally nimble. Initial experimental results show that LeanVec-Sphering and GleanVec push the state of the art for vector search.
\end{abstract}

\section{Introduction}

In recent years, deep learning models have become very adept at creating high-dimensional embedding vectors whose spatial similarities reflect the semantic affinities between its inputs (e.g., images, audio, video, text, genomics, and computer code~\cite{devlin2018bert,radford2021learning,shvetsova2022everything,ji2021dnabert,li2022competition}). This capability has enabled a wide range of applications~\cite[e.g.,][]{blattmann2022retrieval,borgeaud2022improving,karpukhin2020dense,lian2020lightrec,grbovic2016scalable} that search for semantically relevant results over massive collections of vectors by retrieving the nearest neighbors to a given query vector.

Even considering the outstanding recent progress in vector search (a.k.a. similarity search)~\cite[e.g.,][]{malkov2018efficient,fu_fast_2019,andre2021quicker_adc,guo2020accelerating,jayaram2019diskann}, the performance of modern indices significantly degrades as the dimensionality increases~\cite{aguerrebere2023similarity, tepper2023leanvec}.
The predominant example are graph indices~\cite[e.g.,][]{arya1993approximate,malkov2018efficient,fu_fast_2019,jayaram2019diskann} that stand out with their high accuracy and performance for moderate dimensionalities ($D \approx 128$)~\cite{aguerrebere2023similarity} but suffer when dealing with the higher dimensionalities typically produced by deep learning models ($D \approx 512,768$ and upwards)~\cite{tepper2023leanvec}. Additionally, constructing a graph-based index is a time-consuming procedure that also worsens with higher dimensionalities, as its speed is proportional to the search speed. Addressing this performance gap is paramount in a world where vector search deployments overwhelmingly use vectors that stem from deep learning models.

The root cause of this performance degradation is that graph search is bottlenecked by the memory latency of the system, as we are fetching database vectors from memory in a random-like access pattern. In particular, the high memory utilization encountered when running multiple queries in parallel drastically increases the memory latency~\cite{Srinivasan2009cmp} to access each vector and ultimately results in suboptimal search performance. Even masterful placement of prefetching instructions in the software cannot hide the increased latency.

Compressing the vectors appears as a viable idea for alleviating memory pressure. Existing vector compression techniques fall short as they are incompatible with the random pattern~\cite{andre2021quicker_adc,guo2020accelerating} or do not provide sufficient compression~\cite{aguerrebere2023similarity}. Vector compression also becomes a more challenging problem when the statistical distributions of the database and the query vectors differ~\cite{jaiswal2022ood} (here, we say that the queries are out-of-distribution or OOD; when the distributions match, the queries are in-distribution or ID). Unfortunately, this occurs frequently in modern applications, with two prominent examples. The first one is cross-modal querying, i.e., where a user uses a query from one modality to fetch similar elements from a different modality~\cite{radford2021learning,yu2022coca,li2023blip}. For instance, in text2image, text queries are used to retrieve semantically similar images. Second, queries and database vectors can be produced by different models as in question-answering applications~\cite{karpukhin2020dense}.

Recently, LeanVec~\cite{tepper2023leanvec} addressed these problems: it obtained good compression ratios via dimensionality reduction, is fully compatible with the random-access pattern, and addresses the ID and OOD scenarios. However, LeanVec is a computationally lightweight \emph{linear} method, which limits the amount of dimensionality reduction that can be applied without introducing hampering distortions in its accuracy.

Motivated by the power of deep neural networks for nonlinear dimensionality reduction~\cite[e.g.,][]{bank2023autoencoders,kingma2019introduction,bardes2021vicreg,zbontar2021barlow}, we tackle these problems by introducing two new dimensionality reduction algorithms, the linear \textbf{LeanVec-Sphering} and the nonlinear \textbf{Generalized LeanVec (GleanVec)}, that accelerate high-dimensional vector search through a reduction in memory bandwidth utilization.

We present the following contributions:
\begin{itemize}[topsep=0ex,itemsep=0ex,leftmargin=4ex]
    \item \textbf{LeanVec-Sphering} outperforms the other existing ID and OOD linear dimensionality reduction methods, as shown in our experimental results, comes with no hyperparameters, and allows to set the target dimensionality during search (in contrast to setting it during index construction as the existing techniques). Moreover, its training is also efficient, only requiring two singular value decompositions.
    \item \textbf{GleanVec}, with its nimble nonlinear dimensionality reduction, presents further improvements over linear methods in search accuracy while barely impacting its performance.
    In particular, its piecewise linear structure is designed, with the use of LeanVec-Sphering, to efficiently perform similarity computations during a graph search.
    
\end{itemize}

The remainder of this work is organized as follows. 
We formally present our main problem in \cref{sec:problem_statement} along with some preliminary considerations. In \cref{sec:leanvec_closed_form}, we introduce LeanVec-Sphering. We then introduce GleanVec, in \cref{sec:gleanvec} and show its superiority over the existing alternatives with an extensive set of experimental results in \cref{sec:results}. We review the existing literature in \cref{sec:background} and its relation to our work. We provide a few concluding remarks in \cref{sec:conclusions}.

\section{Problem statement}
\label[section]{sec:problem_statement}

\textbf{Notation.} In this work, we denote vectors and matrices by lowercase and uppercase bold letters, respectively, e.g., $\vect{v} \in \Real^{n}$ and $\vect{A} \in \Real^{m \times n}$. We denote sets with a caligraphic font, e.g., $\set{S}$.

We start from a set of database vectors $\set{X} = \left\{ \vect{x}_i \in \Real^D \right\}_{i=1}^n$ to be indexed and searched.
We use maximum inner product as the vector search metric, where one seeks to retrieve for a query $\vect{q}$ the $k$ database vectors with the highest inner product with the query, i.e., a set $\set{N}$ such that
$\set{N} \subseteq \set{X}$, $|\set{N}| = k$, and $(\forall \vect{x}_k \in \set{N}, \forall \vect{x}_i \in \set{X} \setminus \set{N}) \ \langle \vect{q}, \vect{x}_k \rangle \geq \langle \vect{q}, \vect{x}_i \rangle$.
Although maximum inner product is the most popular choice for deep learning vectors, this choice comes without loss of generality as the common cosine similarity and Euclidean distance we can be trivially mapped to this scenario by normalizing the vectors (see \cref{sec:datasets}).

In most practical applications, accuracy is traded for performance to avoid a linear scan of $\set{X}$, by relaxing the definition to allow for a certain degree of error, i.e., a few of the retrieved elements (the approximate nearest neighbors) may not belong to the ground-truth top $k$ neighbors. A vector search index enables sublinear (e.g., logarithmic) searches in $\set{X}$.

Our goal is to accelerate vector search in high-dimensional spaces. The \textbf{baseline search} consists in retrieving a set $\set{N} \subset \{1, \dots, n\}$ of $k$ approximate nearest neighbors to $\vect{q}$ from $\set{X}$ using an index.
Here, graph indices~\cite[e.g.,][]{arya1993approximate,malkov2018efficient,jayaram2019diskann} constitute the state-of-the-art with their high accuracy and performance~\cite{aguerrebere2023similarity}.
Graph indices consist of a directed graph where each vertex corresponds to a database vector and edges represent neighbor-relationships between vectors so that the graph can be efficiently traversed to find the approximate nearest neighbors in sub-linear time.
However, graph indices struggle in high-dimensional spaces~\cite{aguerrebere2023similarity,tepper2023leanvec}.


To address these difficulties, we seek to learn two nonlinear functions $f, g : \Real^D \to \Real^d$ that reduce the dimensionality from $D$ to $d$ such that the inner product between a query $\vect{q}$ and a database vector $\vect{x}$ is preserved, i.e.,
\begin{equation}
    \langle \vect{q}, \vect{x} \rangle \approx \langle f(\vect{q}), g(\vect{x}) \rangle .
    \label[equation]{eq:low_dimensional_approximation}
\end{equation}
Operating with a reduced dimensionality accelerates the inner product by decreasing the amount of fetched memory and accelerating the computation by a factor of $D/d$ at the cost of introducing some errors due to the approximation.

This approximation leads to a decrease in the effective memory bandwidth and the computational pressure when utilized in a \textbf{multi-step search}~\cite{tepper2023leanvec}, as detailed in \cref{algo:search}. This procedure follows the standard search-and-rerank paradigm popular in vector search.
During the \textbf{preprocessing} step (\cref{line:search_preprocessing}), we apply the nonlinear dimensionality reduction to the query $\vect{q}$. This operation is done once per search. We conduct the \textbf{main} search (\cref{line:search_main}) on the low-dimensional vectors using the approximation in \cref{eq:low_dimensional_approximation} to retrieve a set of $\kappa > k$ candidates. Finally, we obtain the final result by re-ranking the candidates according to their similarity in the high-dimensional space, correcting errors due to the approximate inner product.
Our main requirement emerges with this setup: the functions $f$ and $g$ must be chosen so that the multi-step algorithm is faster than the baseline and equally accurate.

\begin{algorithm2e}[t]
    \caption{Multi-step vector search}
    \label{algo:search}

    \KwIn{%
        database vectors $\set{X}_{\text{low}} = \left\{ g(\vect{x}_i) \in \Real^d \right\}_{i=1}^{n}$
        and $\set{X} = \left\{ \vect{x}_i \in \Real^D \right\}_{i=1}^{n}$,
        graph index $\set{G}$,
        query vector $\vect{q} \in \Real^D$,
        the number of $k$ neighbors,
        the number of candidates $\kappa \geq k$.}
    \KwOut{the set $\set{N} \subset \{1, \dots, n\}$ of $k$ approximate nearest neighbors.}

    \textbf{Preprocessing:} apply the nonlinear dimensionality reduction
    $\vect{q}' \gets f(\vect{q})$\;
    \label[line]{line:search_preprocessing}

    \textbf{Main:}
    Retrieve a set $\set{N}_{\text{low}} \subset \{1, \dots, n\}$ of $\kappa$ approximate nearest neighbors to $\vect{q}'$ from $\set{X}_{\text{low}}$ using the graph index $\set{G}$\;
    \label[line]{line:search_main}

    \textbf{Postprocessing:}
    Select the set $\set{N}$ of the top-$k$ elements in $\set{N}_{\text{low}}$, according the inner product between $\vect{q}$ and the corresponding vectors in $\set{X}$, i.e.,
    $(\forall k \in \set{N}, \forall i \in \set{N}_{\text{low}} \setminus \set{N}) \ \langle \vect{q}, \vect{x}_k \rangle \geq \langle \vect{q}, \vect{x}_i \rangle$\;
    \label[line]{line:search_postprocessing}
\end{algorithm2e}

\begin{figure}
    \centering
    \includegraphics[width=0.4\columnwidth]{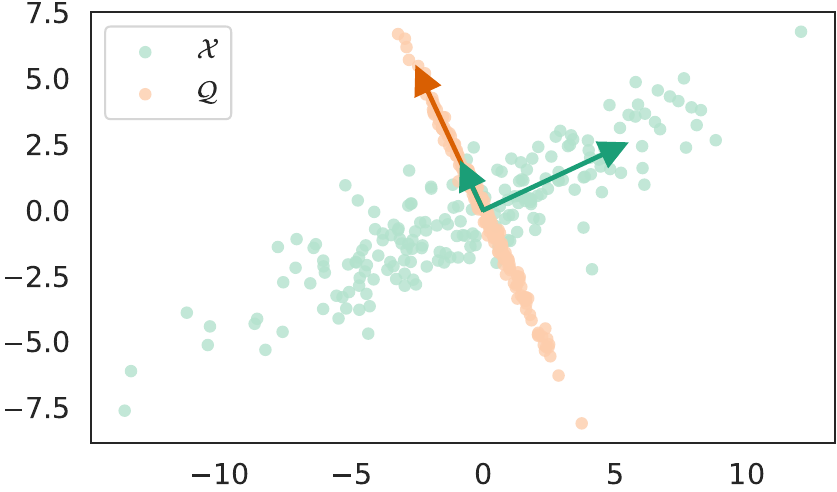}
    \caption{An intuitive cartoon example of the importance of query-aware dimensionality reduction for maximum inner product search. The optimal solution for a query-agnostic method would be to project the database ($\set{X}$) and the query ($\set{Q}$) vectors onto the first principal axis of $\set{X}$ (large green arrow). This choice will yield a poor resolution of the inner products as this direction is orthogonal to the principal axis of $\set{Q}$ (orange arrow), i.e., $(\forall \vect{q} \in \set{Q}, \forall \vect{x} \in \set{X})\ \langle \vect{q}, \vect{x} \rangle \approx 0$. Using a query-aware technique, we can select a direction that maximally preserves the inner products, in this case the second principal direction of $\set{X}$ and the principal direction of $\set{Q}$ coincide and provide the best choice.}
    \label{fig:cartoon_query_aware}
\end{figure}

In modern applications such as cross-modal search (e.g., in text2image~\cite{radford2021learning,yu2022coca,li2023blip}) or cross-model search (i.e., when queries and database vectors are produced by different models as in question-answering~\cite{karpukhin2020dense}), the database and query vectors come from different statistical distributions, which need to be considered simultaneously to successfully reduce the dimensionality of both sets of vectors~\cite{guo2020accelerating,jaiswal2022ood,tepper2023leanvec}.
Thus, we seek to optimize the inner-products directly in a query-aware fashion as in \cref{eq:low_dimensional_approximation}. Not considering the queries and following the more traditional approach of minimizing the reconstruction error, e.g., $\norm{\vect{x} - g^{-1}(g(\vect{x}))}{2}$, and approximating $\langle \vect{q}, \vect{x} \rangle \approx \langle g(\vect{q}), g(\vect{x}) \rangle$, might lead to greater inaccuracies in the inner-product computation, as depicted in \cref{fig:cartoon_query_aware}.

As a possible alternative to \cref{{eq:low_dimensional_approximation}}, we could pose the vanilla approximation $\langle \vect{q}, \vect{x} \rangle \approx \langle \vect{q}, h(g(\vect{x})) \rangle$,
where $f : \Real^D \to \Real^d$ and $h : \Real^d \to \Real^D$ are encoder and decoder functions that reduce and expand the dimensionality, respectively. Here, fetching $g(\vect{x})$ from memory would carry similar memory bandwidth savings as those achieved with \cref{eq:low_dimensional_approximation}. However, the decoder $h$ would need to be applied online before each inner product computation, increasing its cost when compared to the baseline $\langle \vect{q}, \vect{x} \rangle$. On the contrary, $\langle f(\vect{q}), g(\vect{x}) \rangle$ reduces this cost.

\subsection{Preliminaries: The linear case}
\label[section]{sec:leanvec}

LeanVec~\cite{tepper2023leanvec} accelerates multi-step vector search (\cref{algo:search}) for high-dimensional vectors by using, in \cref{eq:low_dimensional_approximation}, the functions
\begin{subequations}
\begin{align}
    \label[equation]{eq:leanvec_approximation_query}
    f(\vect{q}) &= \mat{A} \vect{q} ,\\
    %
    \label[equation]{eq:leanvec_approximation_db}
    g(\vect{x}) &= \mat{B} \vect{x} ,
\end{align}
\end{subequations}
where $\mat{A}, \mat{B} \in \Real^{d \times D}$ are projection matrices, $d < D$.LeanVec is a linear method as the functions $f, g$ are linear.

The projection matrices reduce the number of entries of the database vectors and the quantization reduces the number of bits per entry. The reduced memory footprint decreases the time it takes to fetch each vector from memory. Furthermore, the lower dimensionality alleviates the computational effort (i.e., requiring fewer fused multiply-add operations) of the main search (\cref{line:search_main} in \cref{algo:search}). The overhead incurred with the computation in \cref{line:search_preprocessing} of \cref{algo:search}, i.e., $\vect{q}' \gets \mat{A} \vect{q}$, is negligible when compared to the time taken by the main search (approximately 1\%). 

Given the database vectors $\set{X}$ and a representative set $\set{Q}$ of query vectors, LeanVec~\cite{tepper2023leanvec} learns the projections matrices $\mat{A}, \mat{B}$ that preserve inner products by solving
\begin{equation}
    \min_{\mat{A}, \mat{B} \in \operatorname{St} (D, d)} \sum_{\vect{q} \in \set{Q}} \sum_{\vect{x} \in \set{X}}
    \left( \langle \mat{A} \vect{q}, \mat{B} \vect{x} \rangle - \langle \vect{q} , \vect{x} \rangle \right)^2 ,
    \label[problem]{prob:leanvec}
\end{equation}
where $\operatorname{St} (D, d) = \{ \mat{U} \in \Real^{d \times D} \,|\, \mat{U} \transpose{\mat{U}} = \mat{I} \}$ is the Stiefel manifold (i.e., the set of row-orthonormal matrices).
This problem can be solved efficiently~\cite{tepper2023leanvec} using either block-coordinate descent algorithms or eigenvector search techniques, with different partical and theoretical tradeoffs.

\subsection{Preliminaries: The nonlinear case}

The progress in dimensionality reduction provided by deep learning methods is undeniable: from autoencoders~\cite{bank2023autoencoders} and their variational counterparts~\cite{kingma2019introduction} to constrastive self-supervised learning methods~\cite[e.g.,][]{bardes2021vicreg,zbontar2021barlow}.

We can extend \cref{prob:leanvec} to the nonlinear case by parameterizing the functions $f, g$ as neural networks, i.e.,
\begin{equation}
    \min_{\theta, \omega} \sum_{\vect{q} \in \set{Q}} \sum_{\vect{x} \in \set{X}}
    \left( \langle f_{\theta} (\vect{q}), g_{\omega}(\vect{x}) \rangle - \langle \vect{q} , \vect{x} \rangle \right)^2 ,
    \label[problem]{prob:neuralnet}
\end{equation}
where $\theta, \omega$ are the parameters of the networks $f, g$, respectively.

The outstanding representational power of deep neural networks comes with equally outstanding computational demands. The time to apply $\vect{q}' \gets f(\vect{q})$, \cref{line:search_preprocessing} in \cref{algo:search}, is measured in milliseconds for modern deep neural networks running on state-of-the-art hardware. The latency of the baseline single-step search, defined in \cref{sec:problem_statement}, is measured in microseconds. Thus, the use of modern neural network architectures becomes absolutely prohibitive in our setting.


\section{A closed-form solution for the linear case}
\label{sec:leanvec_closed_form}

The LeanVec loss $\mathscr{L}$ in \cref{prob:leanvec} can be rewritten as
\begin{subequations}
\begin{align}
    \mathscr{L}
    &=
    \sum_{\vect{x} \in \set{X}} \norm{ \transpose{\mat{Q}} \left( \transpose{\mat{A}} \mat{B} \vect{x} - \vect{x} \right) }{2}^2 \\
    &=
    \sum_{\vect{x} \in \set{X}} \traceone{ \transpose{\left( \transpose{\mat{A}} \mat{B} \vect{x} - \vect{x} \right)} \mat{Q} \transpose{\mat{Q}} \left( \transpose{\mat{A}} \mat{B} \vect{x} - \vect{x} \right) } \\
    &=
    \sum_{\vect{x} \in \set{X}} \traceone{ \transpose{\left( \transpose{\mat{A}} \mat{B} \vect{x} - \vect{x} \right)} \mat{U} \mat{S}^2 \transpose{\mat{U}} \left( \transpose{\mat{A}} \mat{B} \vect{x} - \vect{x} \right) } \\
    &=
    \sum_{\vect{x} \in \set{X}} \traceone{ \transpose{\left( \transpose{\mat{A}} \mat{B} \vect{x} - \vect{x} \right)}  \transpose{\mat{W}} \mat{W} \left( \transpose{\mat{A}} \mat{B} \vect{x} - \vect{x} \right) } \\
    &=
    \sum_{\vect{x} \in \set{X}} \norm{ \mat{W} \left( \transpose{\mat{A}} \mat{B} \vect{x} - \vect{x} \right) }{2}^2 ,
    \label[equation]{eq:leanvec_sphering}
\end{align}
\end{subequations}
where $\mat{Q}$ is a $D \times m$ matrix obtained by stacking the vectors in $\set{Q}$ horizontally, $\mat{U} \mat{S} \transpose{\mat{V}}$ is the singular value of $\mat{Q}$, and $\mat{W} = \mat{U} \mat{S} \transpose{\mat{U}}$. The derivation emerges from the  identities
\begin{equation}
    \mat{Q} \transpose{\mat{Q}}
    =
    \mat{U} \mat{S} \transpose{\mat{V}} \mat{V} \mat{S} \transpose{\mat{U}}
    =
    \mat{U} \mat{S}^2 \transpose{\mat{U}}
    =
    \mat{U} \mat{S} \transpose{\mat{U}} \mat{U} \mat{S} \transpose{\mat{U}}
    =
    \transpose{\mat{W}} \mat{W}
    \label[equation]{eq:QQt_identitites}
\end{equation}
by virtue of $\mat{U}$ and $\mat{V}$ being unitary matrices.

From \cref{eq:leanvec_sphering}, it is now clear that we are actually trying to find projection matrices $\mat{A}$ and $\mat{B}$ that reduce the dimensionality under a Mahalanobis distance with weight matrix $\mat{W}$. This can be interpreted as using the Euclidean distance after a whitening or sphering transformation, with the difference that here the matrix $\mat{W}$ is computed from $\set{Q}$ instead of $\set{X}$ as in the classical whitening transformation.

In this work, we restrict the forms of $\mat{A}$ and $\mat{B}$ to $\mat{A} = \mat{P} \mat{W}^{-1}$ and $\mat{B} = \mat{P} \mat{W}$ where $\mat{P} \in \operatorname{St} (D, d)$ is an unknown matrix.
We assume for simplicity that $\mat{W}$ is invertible; if not, we can use a pseudoinverse. These restrictions lead to the new problem
\begin{align}
    \min_{\mat{P} \in \operatorname{St} (D, d)}
    \sum_{\vect{x} \in \set{X}} \norm{ \transpose{\mat{P}} \mat{P} \mat{W} \vect{x} - \mat{W} \vect{x} }{2}^2 .
    \label[problem]{eq:leanvec2_sphering}
\end{align}
The optimization of \cref{eq:leanvec2_sphering} boils down to a singular value decomposition (SVD) of the matrix $\mat{W} \mat{X}$, where $\mat{X}$ is obtained by stacking the vectors in $\set{X}$ horizontally.

In our model, we have
\begin{equation}
    \langle \vect{q}, \vect{x} \rangle
    \approx \langle \vect{q}, \mat{W}^{-1} \transpose{\mat{P}} \mat{P} \mat{W} \vect{x} \rangle
    = \langle \mat{P} \mat{W}^{-1} \vect{q}, \mat{P} \mat{W} \vect{x} \rangle
    = \langle \mat{A} \vect{q}, \mat{B} \vect{x} \rangle ,
\end{equation}
The application of $\mat{W}^{-1}$ to the query vectors flattens the spectrum of their distribution, which becomes ``spherical.'' Thus, we refer to this new approach as LeanVec-Sphering and summarize it in \cref{algo:leanvec_sphering}.
Critically, this algorithm does not involve any hyperparameters beyond the target dimensionality $d$.
Finally, we could apply scalar quantization to the database vectors $\mat{B} \vect{x}$ with \cref{eq:leanvec_approximation_db} as in LeanVec.

\setlength{\algomargin}{1em}
\begin{algorithm2e}[t]    
    \caption{LeanVec-Sphering algorithm.}
    \label[algorithm]{algo:leanvec_sphering}

    \KwIn{%
        $D$-dimensional learning database vectors $\set{X}_{\text{learn}}$ and
        learning query vectors $\set{Q}_{\text{learn}}$,
        the target dimensionality $d \in [1, D]$.}
    \KwOut{projection matrices $\mat{A}$ and $\mat{B}$.}

    Form $\mat{Q}$ and $\mat{X}$ by horizontally stacking the vectors in $\set{Q}_{\text{learn}}$ and in $\set{X}_{\text{learn}}$, respectively.

    Compute the SVD $\mat{U} \mat{S} \transpose{\mat{V}}$ of $\mat{Q}$\;
    \label[line]{algo_line:svd_queries}

    $\mat{W} \gets \mat{U} \mat{S} \transpose{\mat{U}}$\;

    Form $\mat{P}$ with the $d$ left singular vectors of $\mat{W} \mat{X}$ corresponding to its largest singular values\;
    \label[line]{algo_line:svd_data}
    
    $\mat{A} \gets \mat{P} \mat{W}^{-1}$\;
    $\mat{B} \gets \mat{P} \mat{W}$\;
\end{algorithm2e}

\textbf{Efficiency.}
The computation of the two SVD applications in \cref{algo:leanvec_sphering} requires $O(D^2 + n D^2 + D^3)$ floating point operations. This makes \cref{algo:leanvec_sphering} only linear in $n$ and cubid in $D$. This has the same algorithmic complexity profile as the original LeanVec algorithms~\cite{tepper2023leanvec}. Moreover, as the sample size grows, the SVD computations converge quickly (at a square-root rate) to their expected value~\cite{koltchinskii2017concentration}. Thus, operating on $\set{X}$ (resp.~$\set{Q}$), on a subsample of $\set{X}$ (resp.~$\set{Q}$), or on a small holdout set $\set{X}_{\text{learn}}$ (resp.~$\set{Q}_{\text{learn}}$) will have a small effect on the end result.

\textbf{Quality.}
In \cref{sec:results}, we show the superiority of the LeanVec-Sphering algorithm over the original LeanVec algorithms~\cite{tepper2023leanvec}. First, LeanVec-Sphering achieves better minima of the loss in \cref{prob:leanvec}, which accounts for all inner products in the training sets. We also analyze the more practically relevant results for maximum inner product search between validation query and database vector sets, where LeanVec-Sphering prevails.

\subsection{Flexibly selecting the target dimensionality}
\label[section]{sec:flexible_target_dimensionality}

\begin{figure}
    \centering
    \includegraphics[width=0.4\columnwidth]{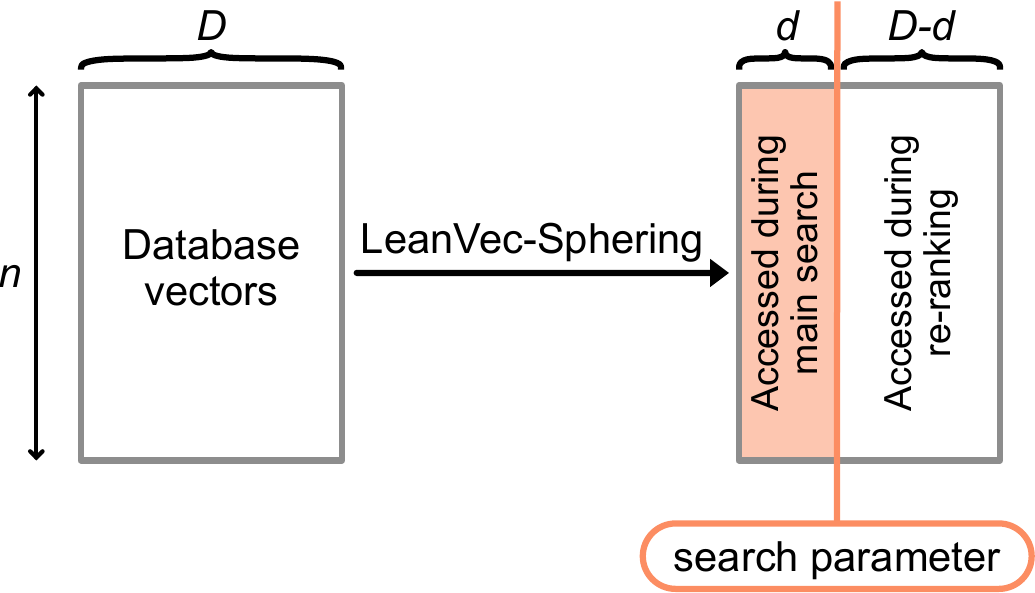}
    \caption{LeanVec-Sphering is a novel query-aware technique for dimensionality reduction from $D$ dimensions to $d<D$ target dimensions. The linear transformation in LeanVec-Sphering allows to select $d$ flexibly during search, instead of fixing its value during the construction of the search index (see \cref{sec:flexible_target_dimensionality} for further details). This enables tuning $d$ when using the index operationally.}
    \label{fig:leanvec_sphering}
\end{figure}

LeanVec-Sphering provides the following mechanism to select the target dimensionality $d$ based on the magnitude of the singular values of $\mat{W} \mat{X}$. We can store $D$-dimensional vectors, i.e.,
\begin{equation}
    \left\{ \vect{x}' = \mat{P}' \mat{W} \vect{x} \,|\, \vect{x} \in \set{X} \right\} .
    \label[equation]{eq:flexible_vector_storage}
\end{equation}
where $\mat{P}' \in \Real^{D \times D}$ contains the left singular vectors of $\mat{W} \mat{X}$ sorted in decreasing order of the singular value magnitudes.
Alternatively, we can store $\alpha d < D$ dimensions for some $\alpha > 1$. With this, we can select the value of $d$ during search by considering the first $d$ dimensions of $\vect{x}'$ (see \cref{fig:leanvec_sphering}) and of $\mat{P}' \mat{W}^{-1} \vect{q}$ when performing their inner product. 

For postprocessing in \cref{algo:search}, we can directly use $\vect{x}'$ without the need to store a set of secondary database vectors as in \cite{aguerrebere2023similarity,tepper2023leanvec}.
We just set $\mat{A}' = \mat{P}' \mat{W}^{-1}$ and $\mat{B}' = \mat{P}' \mat{W}$ and obtain
\begin{equation}
    \langle \vect{q}, \vect{x} \rangle
    = \langle \mat{A}' \vect{q}, \mat{B}' \vect{x} \rangle
    = \langle \mat{A}' \vect{q}, \vect{x}' \rangle .
\end{equation}
The equality allows to compute undistorted similarities.
Furthermore, this enables handling queries whose distribution differs from the one of the set used to build the projection matrices. For example, in a text-to-image application, we may be also interested in running image queries as a secondary option.

The approach described in the previous paragraph may entail some performance penalties, as memory pages misses may be increased as a consequence of storing all $D$ dimensions for each vector instead of $d$ (thus storing fewer vectors per page). However, preliminary experiments show that this is a second-order effect, largely overshadowed by the increase in usability.

Effectively, $d$ has become a tunable search hyperparameter that can be used to tradeoff accuracy for performance and viceversa seamlessly at runtime without changing the underlying index.
This flexibility is a remarkable improvement over LeanVec~\cite{tepper2023leanvec}, where $d$ needed to be selected following a trail-and-error approach during the construction of the search index (albeit choosing a conservative value was indeed possible). The Frank-Wolfe and the eigensearch algorithms in LeanVec~\cite{tepper2023leanvec} will yield different matrices $\mat{A}$ for different values of $d$ that are not related to each other by their spectrum (and similarly for $\mat{B}$).

\subsection{Streaming vector search}

In streaming vector search, the database is built dynamically by adding and removing vectors.
We start from an initial database $\set{X}_{0} = \{ \vect{x} \in \Real^D \}$, containing $n_0$ vectors in $D$ dimensions. Then, at each time $t > 1$, a vector $\vect{x}_i$ is either inserted (i.e., $\set{X}_{t} = \set{X}_{t - 1} \cup \{ \vect{x}_i \}$) or removed (i.e., $\set{X}_{t} = \set{X}_{t - 1} \setminus \{ \vect{x}_i \}$). We assume with no loss of generality that each vector has a universally unique ID.
There are two main challenges for linear dimensionality reduction in this setting.

First, we need to keep the projection matrices $\mat{A}$ and $\mat{B}$ up to date for each $t$.
We assume that the set $\set{Q}_{t}$ of query vector also evolves over time.
We can do this by keeping the $D \times D$ summary statistics
\begin{equation}
    \mat{K}_{\set{Q}_{t}} = \sum_{\vect{q} \in \set{Q}_{t}} \vect{q} \transpose{\vect{q}}
    \quad\text{and}\quad
    \mat{K}_{\set{X}_{t}} = \sum_{\vect{x} \in \set{X}_{t}} \vect{x} \transpose{\vect{x}} .
\end{equation}
The matrix $\mat{K}_{\set{X}_{t}}$ can be directly updated when inserting/removing a vector $\vect{x}_i$ by adding/subtracting its outer product $\vect{x}_i \transpose{\vect{x}_i}$ (and similarly for $\mat{K}_{\set{Q}_{t}}$). We can compute $\mat{A}$ and $\mat{B}$ by replacing the SVD computations in \cref{algo_line:svd_queries,algo_line:svd_data} in \cref{algo:leanvec_sphering} by the eigendecompositions of $\mat{K}_{\set{Q}_{t}}$ and of $\mat{W} \mat{K}_{\set{X}_{t}} \mat{W}$, respectively. This equivalence follows directly from the equivalence between the SVD of a matrix $\mat{M}$ and the eigendecomposition of $\mat{M} \transpose{\mat{M}}$.

Second, we need to be able to reapply the matrix $\mat{B}$ to the database vectors $\set{X}$ without holding a copy of $\set{X}$ in memory (or materializing it again from disk).
An immediate solution is to recompute $\mat{W}$ and $\mat{P}$ every $s$ temporal updates, i.e., when $s \lfloor t / \rfloor s = t$.
Following \cref{sec:flexible_target_dimensionality}, we can store the set $\{ \mat{P}_{t'} \mat{W}_{t'} \vect{x} \,|\, \vect{x} \in \set{X}_{t} \}$ of database vectors in memory, where $t' = s \lfloor t / s \rfloor$. To update the existing database vectors upon computing $\mat{W}_{t'+1}$, we can undo the application of $\mat{W}_{t'}$ and apply $\mat{W}_{t'+1}$, i.e.,
\begin{equation}
    \left\{ \mat{P}_{t'+1} \mat{W}_{t'+1} \left( \mat{P}_{t'} \mat{W}_{t'} \right)^{-1} \mat{P}_{t'} \mat{W}_{t'} \vect{x} \,|\, \vect{x} \in \left( \set{X}_{t} \cap \set{X}_{t+1} \right) \right\} .
\end{equation}
These updates can be applied sequentially by iterating over the previously existing elements, $\set{X}_{t} \cap \set{X}_{t+1}$, or on-demand when one of these vectors is accessed during search.
For the newly introduced vectors, i.e., those in $\set{X}_{t+1} \setminus \set{X}_{t}$, we have
\begin{equation}
    \left\{ \mat{P}_{t'+1} \mat{W}_{t'+1} \vect{x} \,|\, \vect{x} \in \left( \set{X}_{t+1} \setminus \set{X}_{t} \right) \right\} .
\end{equation}

\section{GleanVec: Generalizing LeanVec with nonlinearities}
\label[section]{sec:gleanvec}

In this work, we take inspiration from different ideas for learning of a nonlinear dimensionality reduction method that is computationally nimble. Our main sources of inspiration are:
\begin{description}[topsep=0ex,itemsep=0ex,leftmargin=4ex]
\item[Inspiration 1.]
Principal Component Analysis (PCA) is the problem of fitting a linear subspace $S \subset \Real^D$ of unknown dimension $d < D$ to $n$ noisy points generated from $S$. Generalized PCA~\cite{vidal2005generalized} considers the extension of PCA to the case of data lying in a union of subspaces, i.e., the combined problem of partitioning the data and reducing the dimensionality in each partition linearly. Generalized LeanVec inherits its name from this connection.

\item[Inspiration 2.]
In the pre-deep-learning era, there is a long history of successfully using k-means clustering as a method to learn features from data~\cite[and references therein]{o2011introduction}. When used as a deep learning technique, k-means has been shown to be competitive with feedforward neural networks~\cite{coates2012learning}. k-means divides the input space following a Voronoi diagram~\cite[e.g.,][]{du1999centroidal}.

More recently, deep neural networks were shown to provide piecewise linear representations~\cite{balestriero2018spline}.
Their layers partition their input space using a Voronoi-like diagram and apply, in each cell, a linear mapping to their input to produce their output~\cite{balestriero2019geometry}.
\end{description}

From these considerations, the core idea behind GleanVec is to decouple the problem in two stages: (1) data partitioning, and (2) locally-linear dimensionality reduction applied in a query-aware fashion. This scheme allows to replace the compute-bound dimensionality reduction from neural networks by a nimble selection-based technique. Note the contrast with CCST~\cite{zhang_connecting_2022} that uses powerful transformers that preclude its usage for search.

For now, we assume that we are given a set of dimensionality-reducing functions $\{ f_c : \Real^D \to \Real^d \}_{c=1}^{C}$ and $\{ g_c : \Real^D \to \Real^d \}_{c=1}^{C}$ that will be applied to the query and database vectors, respectively, and a set of landmarks $\{ \vect{\mu}_{c} \in \Real^D \}_{c=1}^{C}$ such that $\norm{\vect{\mu}_{c}}{2} = 1$.
To each vector $\vect{x}_i \in \set{X}$, we associate  the tag $c_i \in \{1, \dots, C\}$ such that
\begin{equation}
    c_i = \argmax_{c \in \{1, \dots, C\}} \langle \vect{x}_i \,,\, \vect{\mu}_{c} \rangle .
    \label[equation]{eq:cluster_tag}
\end{equation}
We keep the tags $\{ c_i \}_{i=1}^n$ in memory.
We also compute the dataset
\begin{equation}
    \set{X}_{\text{low}} = \left\{ \vect{x}^{\text{low}}_i \right\}_{i=1}^{n} ,
    \quad\text{where}\quad
    \vect{x}^{\text{low}}_i = f_{c_i} (\vect{x}_i) .
    \label[equation]{eq:X_low}
\end{equation}
With this setup, we propose to approximate the inner product by
\begin{equation}
    \langle \vect{q}, \vect{x}_i \rangle
    \approx \langle f_{c_i}(\vect{q}), g_{c_i}(\vect{x}) \rangle 
    = \langle f_{c_i}(\vect{q}), \vect{x}^{\text{low}}_i \rangle .
    \label[equation]{eq:low_dimensional_approximation_gleanvec}
\end{equation}
Here, we replaced the application of a computationally-complex nonlinear function $f(\vect{q})$ by the selection of a computationally simple function $f_{c_i}$. This selection only requires one memory lookup to retrieve the tag $c_i$ associated with $\vect{x}_i$ via \cref{eq:cluster_tag}. This lookup can be efficiently performed by storing the pair $(c_i, \vect{x}^{\text{low}}_i)$ contiguously in memory.

By carefully choosing computationally efficient functions $f_{c}$, we can accurately approximate and accelerate the inner product computation. To achieve this simplicity, we opt to use linear functions $f_{c}$ and $g_{c}$, i.e.,
\begin{subequations}
\begin{align}
    f_c(\vect{q}) &= \mat{A}_c \vect{q} ,\\
    g_c(\vect{x}) &= \mat{B}_c \vect{x} .
\end{align}
\end{subequations}
These function are learned from data in a query-aware fashion as detailed in \cref{sec:gleanvec_learning}.

Each time we need to compute an inner product using \cref{eq:low_dimensional_approximation_gleanvec} during the main search, we can resort to two different techniques: (1) lazy execution (\cref{algo:gleanvec_ip_lazy}), where the matrix-vector multiplications are performed on-the-fly; or (2) eager execution (\cref{algo:gleanvec_ip_eager}), where we precompute the different low-dimensional views of the query, $(\forall c = 1, \dots, C)\, \vect{q}_c = \mat{A}_c \vect{q}$, once and select the appropriate one when computing each inner product. A similar eager computation is used for scalar quantization in \cite{aguerrebere2024high}. The lazy alternative involves more floating-point operations as, during each search, we encounter multiple vectors that belong to the same cluster $c$ and we thus repeat the operation $\mat{A}_c \vect{q}$ multiple times. The eager alternative, on the other side, might compute some vectors $\vect{q}_c$ that are never used throughout the search. The efficiency of the eager algorithm mainly depends on the success of the system in retaining the vectors $\vect{q}_c$ in higher levels of the cache hierarchy. Ultimately, the choice comes down to an empirical validation that we explore in \cref{sec:results}.

\begin{algorithm2e}[t]
    \caption{Lazy inner product with GleanVec}
    \label{algo:gleanvec_ip_lazy}

    Recover the assignment $c_i$ for $\vect{x}_i$\;
    \Return $\langle \mat{A}_{c_i} \vect{q} \,,\, \vect{x}^{\text{low}}_i \rangle$%
    \tcp*{$\vect{x}^{\text{low}}_i$ defined in Eq.~(\ref{eq:X_low})}
\end{algorithm2e}

\begin{algorithm2e}[t]
    \caption{Eager inner product with GleanVec}
    \label{algo:gleanvec_ip_eager}

    \SetKwProg{myproc}{Procedure}{}{}

    \SetKwFunction{preprocess}{preprocess}
    \myproc{\preprocess{$\vect{q}$}}{
        \For{$c = 1, \dots, C$}{
            $\vect{q}_c \gets \mat{A}_c \vect{q}$\;
        }
        
    }

    \SetKwFunction{distance}{inner-product}
    \myproc{\distance{$\vect{x}_i$, $\vect{q}$}}{
        Recover the assignment $c_i$ for $\vect{x}_i$\;
        \Return $\langle \vect{q}_{c_i} \,,\, \vect{x}^{\text{low}}_i \rangle$%
        \tcp*{$\vect{x}^{\text{low}}_i$ defined in Eq.~(\ref{eq:X_low})}
    }
\end{algorithm2e}

\subsection{Learning}
\label[section]{sec:gleanvec_learning}

During learning, we use learning sets $\set{X}_{\text{learn}}$ and $\set{Q}_{\text{learn}}$ which may be subsets of $\set{X}$ and $\set{Q}$, respectively, or holdout sets. 

First, we partition $\set{X}_{\text{learn}}$ into $C$ subsets, i.e., $\set{X}_{\text{learn}} = \bigcup_{c=1}^C \set{X}_c$ and $\set{X}_c \cap \set{X}_{c'} = \emptyset$ for all $c \neq c'$. Then, for each subset $\set{X}_c$, we seek a pair of linear functions $f_c$ and $g_c$ such that
\begin{equation}
    (\forall \vect{q} \in \set{Q}_{\text{learn}} ,\, \forall \vect{x} \in \set{X}_c)\ 
    \langle \vect{q}, \vect{x} \rangle \approx \langle f_c(\vect{q}) , g_c(\vect{x})  \rangle ,
\end{equation}
where
$f_c(\vect{q}) = \mat{A}_c \vect{q}$ and
$g_c(\vect{x}) = \mat{B}_c \vect{x}$.

\begin{algorithm2e}[t]
    \caption{GleanVec optimization}
    \label{algo:glean_vec_learning}

    \KwIn{%
        learning query and database sets, $\set{Q}_{\text{learn}}$ and $\set{X}_{\text{learn}}$ with $D$-dimensional vectors,
        the number of clusters $C$,
        the target dimensionality $d \in [1, D]$.}
    \KwOut{cluster centers $\{ \vect{\mu}_c \}_{c=1}^C$, projection matrices $\left\{ \mat{A}_c \in \Real^{d \times D} \right\}_{c=1}^C$ and $\left\{ \mat{B}_c \in \Real^{d \times D} \right\}_{c=1}^C$.}

    Let $\widetilde{\set{X}} \gets \left\{ \vect{x}_i / \norm{\vect{x}_i}{2} \,|\, \vect{x} \in \set{X}_{\text{learn}} \right\}$\;
    Learn $C$ landmarks $\vect{\mu}_c$ from $\widetilde{\set{X}}$ with spherical k-means\;
    \For(\tcp*[h]{in parallel}){$c=1, \dots, C$}{
        Let $\set{X}_c = \left\{ \vect{x}_i \in \vect{X} \,|\, c = \argmax_{c'} \langle \vect{x}_i / \norm{\vect{x}_i}{2} \,,\,  \vect{\mu}_{c'} \rangle \right\}$\;
        Compute the projection matrices $\mat{A}_c, \mat{B}_c$ from $\set{Q}_{\text{learn}}$ and $\set{X}_c$ with \cref{algo:leanvec_sphering}\;
    }
\end{algorithm2e}

We propose \cref{algo:glean_vec_learning} to effectively learn the different GleanVec parameters. There, we perform the partitioning with spherical k-means~\cite{dhillon2001concept} (more details in \cref{sec:spherical_kmeans}) to learn cluster centers $\{ \vect{\mu}_c \}_{c=1}^C$. Then,
\begin{equation}
    \set{X}_c = \left\{ \vect{x} \in \set{X}_{\text{learn}} \ \Big|\ c = \argmax_{c' \in \{1, \dots, C\}} \langle \vect{x} \,,\, \vect{\mu}_{c'} \rangle \right\} .
\end{equation}
The choice of the clustering technique, is guided by two main principles. Critically, we need a model that allows assigning cluster tags to data not seen during learning as this will allow to insert new vectors into our vector search index later on. Additionally, we also want this inference task to be fast to quickly process these insertions.
Finally, given $\set{Q}_{\text{learn}}$ and $\set{X}_c$, we reduce the dimensionality using the proposed LeanVec-Sphering described in \cref{sec:leanvec_closed_form}.



\subsection{An alternative perspective on GleanVec}

GleanVec can be regarded as a new type of neural network that borrows elements from the Vector Quantized Variational Autoencoder (VQ-VAE)~\cite{van2017neural}. While VQ-VAE uses cluster centers to perform vector quantization, GleanVec uses them for dimensionality reduction.

In VQ-VAE, the intermediary input $\vect{x}$ is passed through a discretization bottleneck followed by mapping onto the nearest center, yielding the intermediary output $z(\vect{x})$, i.e.,
\begin{equation}
    z(\vect{x}) = \vect{\mu}_{c^*},
    \quad\text{where}\quad
    c^* = \argmin_{c \in \{1, \dots, C\}} \norm{ \vect{x}_i - \vect{\mu}_{c} }{2} .
\end{equation}
This bottleneck layer is surrounded upstream and downstream by other neural network layers (e.g., CNNs).

\begin{figure}
    \centering
    \includegraphics[width=0.55\textwidth]{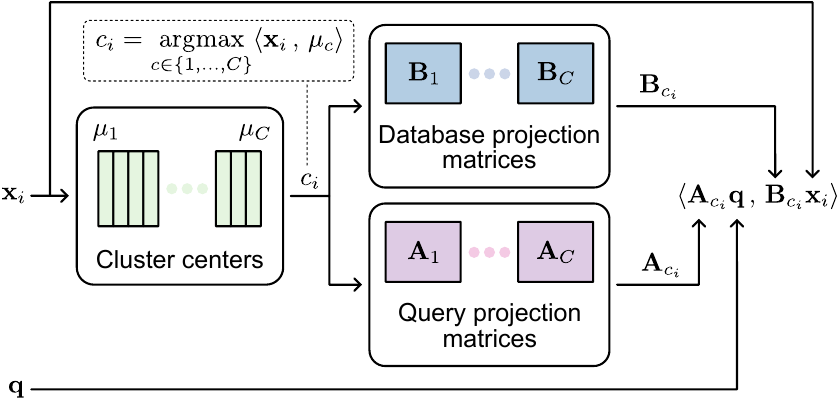}
    \caption{GleanVec can be seen as a new type of neural network that borrows elements from the Vector Quantized Variational Autoencoder (VQ-VAE)~\cite{van2017neural}. While VQ-VAE uses cluster centers to perform vector quantization, GleanVec uses them for dimensionality reduction.}
    \label{fig:gleanvec_neuralnet}
\end{figure}

In GleanVec, the input $\vect{x}$ is passed through a discretization bottleneck followed by linear mappings to $\vect{q}$ and $\vect{c}$, yielding the outputs $z_q(\vect{q})$ and $z_x(\vect{x})$, i.e.,
\begin{equation}
    \begin{aligned}
        z_q(\vect{q}) = \mat{A}_{c_i} \vect{q}, \\   
        z_x(\vect{x}) = \mat{B}_{c_i} \vect{x},    
    \end{aligned}
    \quad\text{where}\quad
    c_i = \argmin_{c \in \{1, \dots, C\}} \norm{ \vect{x}_i - \vect{\mu}_{c} }{2} .
\end{equation}
This computational block is depicted in \cref{fig:gleanvec_neuralnet}.

Interestingly, this perspective sheds light on the conceptual differences between GleanVec and the nowadays standard neural network approach in \cref{prob:neuralnet}. In the latter, the networks $f_{\theta}$ and $g_{\omega}$ only interact through the loss function whereas in GleanVec the output $f(\vect{q}) = z_q(\vect{q})$ directly depends on $\vect{x}_i$. This increases the locality of the formulation and its representational power. Additionally, we could replace the linear functions $f_c, g_c$ in GleanVec by nonlinear functions. This alternative is left for the future work.

GleanVec could be learned in an end-to-end fashion as VQ-VAE, using appropriate techniques to propagate the gradient through the bottleneck~\cite{van2017neural,vali2022nsvq}.
From this perspective the learning algorithm presented in \cref{sec:gleanvec_learning} can be regarded as a stage-wise training technique. We leave for the future the exploration of this end-to-end learning technique.

\section{Experimental results}
\label[section]{sec:results}


\begin{table}[t]
    \centering
    \setlength\extrarowheight{2pt}
    \caption{Evaluated datasets, where $n$ is the number of database vectors (in millions) and $D$ their dimensionality. In all cases, we select the target dimensionality $d$ that yields maximum performance at 90\% accuracy (10-recall@10). The datasets are originally encoded using 32-bits floating-point values. We use separate learning and test query sets, each with 10K entries. We consider the inner product (IP) similarity (see \cref{sec:datasets} for further details, particularly pertaining $D=961$ in GIST-1M-ID).}
    \label[table]{table:datasets}
    
    
    \begin{minipage}{0.48\columnwidth}
    \small
    \centering
    \begin{tblr}
        {
            colsep=2pt,
            colspec={l c S[table-format=3] S[table-format=3]}
        }
    
        \hline
        \SetCell[c=4]{c} In-distribution \\
        \hline
        \SetCell[c=1]{c} Dataset & {{{$n$}}} & {{{$D$}}} & {{{$d$}}}\\
        
        \hline
        GIST-1M-ID & 1M   & 961 & 160\\
        DEEP-1M-ID & 1M   & 256 & 96\\
        LAION-1M-ID & 1M  & 512 &  320 \\
        OI-1M-ID & 1M  & 512 & 160\\
        OI-13M-ID & 13M  & 512 & 160\\
        RQA-1M-ID & 1M    & 768 &  160 \\
        \hline
    \end{tblr}
    \end{minipage}%
    \hfill%
    \begin{minipage}{0.5\columnwidth}
    \small
    \centering
    \begin{tblr}
        {
            colsep=2pt,
            colspec={l c S[table-format=3] S[table-format=3]}
        }
    
        \hline
        \SetCell[c=4]{c} Out-of-distribution \\
        \hline
        \SetCell[c=1]{c} Dataset & {{{$n$}}} & {{{$D$}}} & {{{$d$}}}\\
        \hline
        T2I-1M-OOD & 1M  & 200 & 192\\
        T2I-10M-OOD & 10M  & 200 & 192\\
        WIT-1M-OOD & 10M   & 512 & 256 \\        
        LAION-1M-OOD & 1M  & 512 &  320 \\
        RQA-1M-OOD & 1M    & 768 &  160 \\
        RQA-10M-OOD & 10M    & 768 & 160 \\
        \hline
    \end{tblr}
    \end{minipage}%

\end{table}

\textbf{Datasets.}
We evaluate the effectiveness of our method on a wide range of datasets with varied sizes ($n=1\text{M}$ to $n=13\text{M}$) and medium to high dimensionalities ($D=200$ to $D=960$), containing in-distribution (ID) and out-of-distribution (OOD) queries, see \cref{table:datasets}. For ID and OOD evaluations, we use standard and recently introduced datasets~\cite{zhang_connecting_2022,text2image,Schuhmann2021,aguerrebere2024high,tepper2023leanvec}. See \cref{sec:datasets} for more details.

\textbf{Metrics}
Search accuracy is measured by $K$-recall$@k$, defined by $| S \cap G_t | / K$, where $S$ are the IDs of the $k$ retrieved neighbors and $G_t$ is the ground-truth. In general, we use the default value $K=10$
Search performance is measured by queries per second (QPS).


\subsection{Linear dimensionality reduction}

We begin by evaluating the proposed LeanVec-Sphering (\cref{sec:leanvec_closed_form}) against the state-of-the-art LeanVec-FW and LeanVec-ES algorithms~\cite{tepper2023leanvec} that optimize \cref{prob:leanvec}. LeanVec-FW is based on a block-coordinate descent algorithm that relies on Frank-Wolfe~\cite{frank1956algorithm} optimization to solve its subproblems. LeanVec-ES uses a binary search technique to find a joint subspace that simultaneously covers the query and database vectors. We also consider a variant (LeanVec-ES+FW) that is initialized with LeanVec-ES and refined with LeanVec-FW.

\begin{figure*}
    \centering
    \includegraphics[width=0.9\textwidth]{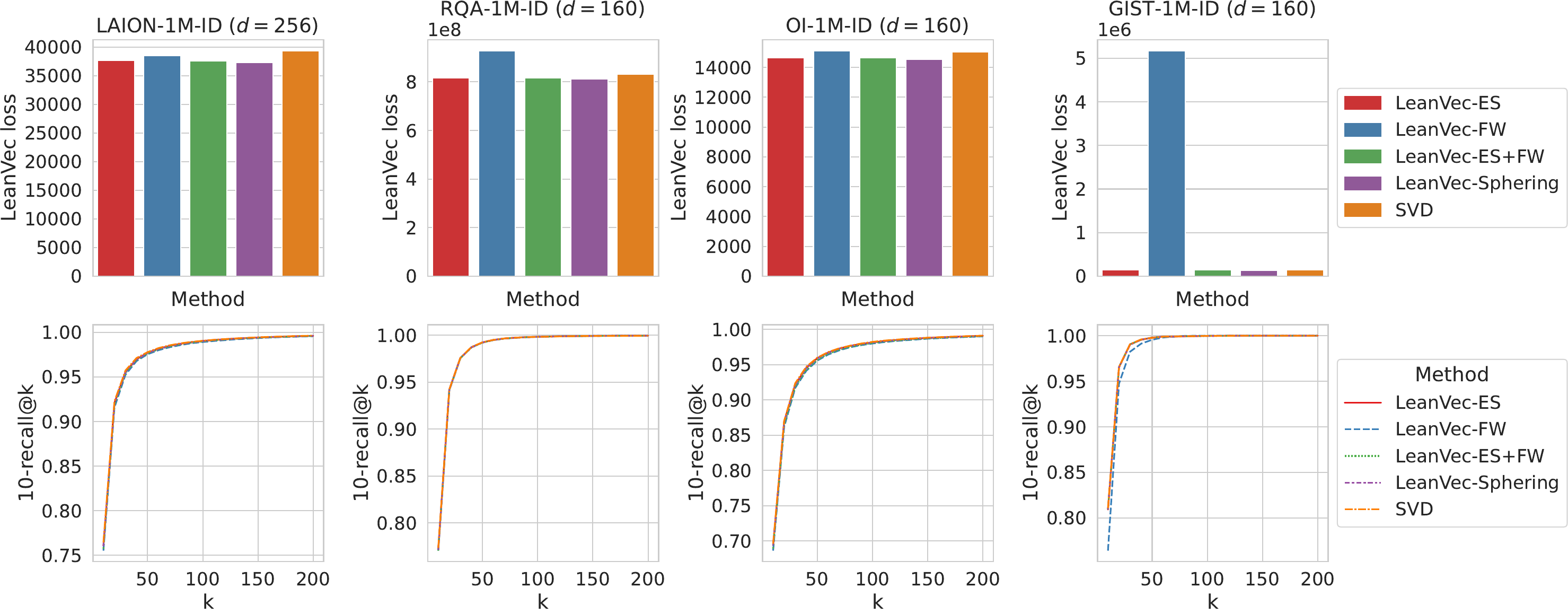}%
    
    \caption{For different ID datasets, all the analyzed methods (including LeanVec-Sphering) perform similarly. This means that using LeanVec-Sphering in the ID setting is safe as it is equivalent to performing a query-agnostic dimensionality reduction with the SVD.}
    \label{fig:comparison_leanvec_ID}
\end{figure*}

In the ID setting of \cref{fig:comparison_leanvec_ID}, all algorithms perform similarly in terms of obtaining comparable values of the loss in \cref{prob:leanvec} (top row) and in search accuracy (bottom row). Although LeanVec-FW achieves a much worse loss in GIST-1M-ID, its search accuracy is comparable to the others. Not surprisingly, a simple SVD succeeds in this case.

\begin{figure*}
    \centering
    \includegraphics[width=0.9\textwidth]{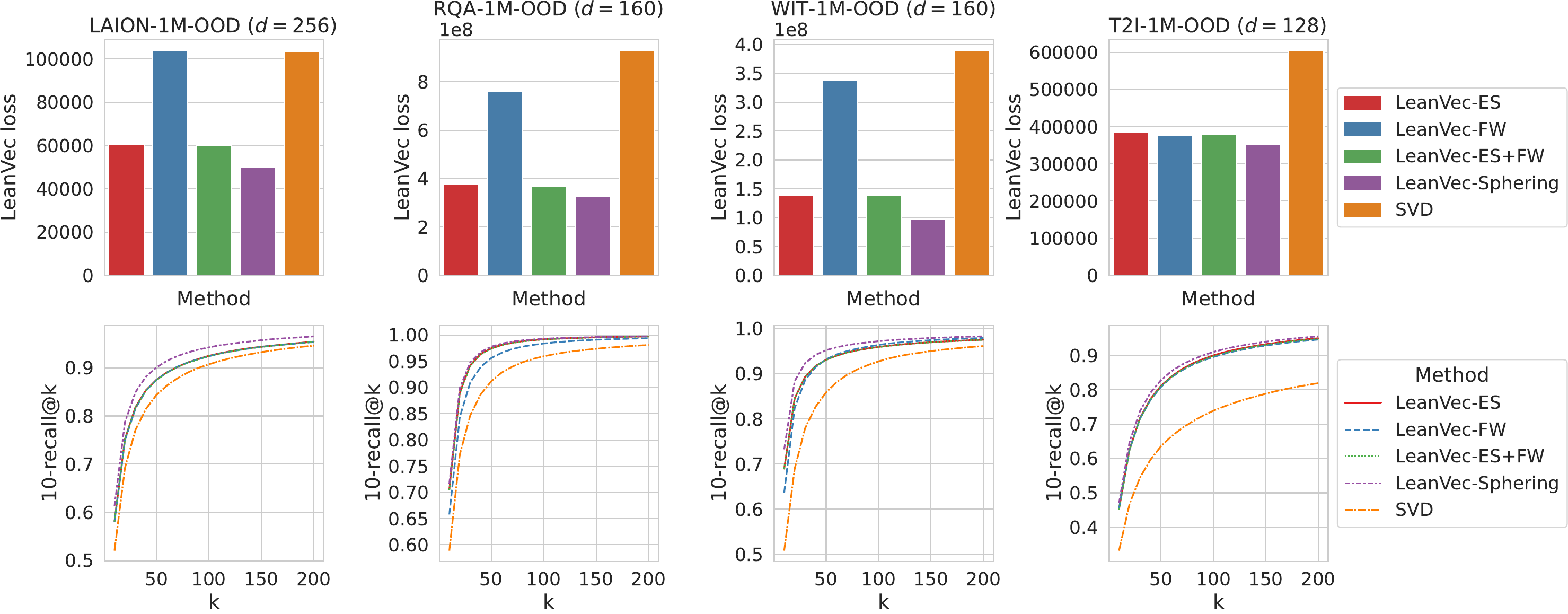}%
    
    \caption{The proposed LeanVec-Sphering outperforms the alternatives by yielding a lower value of LeanVec loss function (top row) and more importantly a higher accuracy for a brute-force search (bottom row) for different OOD datasets and target dimensionalities).}
    \label{fig:comparison_leanvec_OOD}
\end{figure*}

The OOD setting of \cref{fig:comparison_leanvec_OOD} paints a different picture. Here, the SVD, that does not consider the distribution of the queries, clearly underperforms. In all cases, LeanVec-Sphering outperforms the alternatives both in loss and search accuracy. The LeanVec-Sphering results, when combined with its increased computational efficiency compared to the other algorithms, make a clear case for its usage across the board.

Note that the computational complexities of all of these linear methods are equivalent (i.e., one matrix multiplication) and thus the improved accuracy of LeanVec-Sphering translates directly to improved search speeds as well. We will include these experiments in an upcoming version of this manuscript.

\subsection{Nonlinear dimensionality reduction}

\begin{figure*}
    \centering
    \begin{minipage}{0.65\textwidth}
        \includegraphics[width=\textwidth]{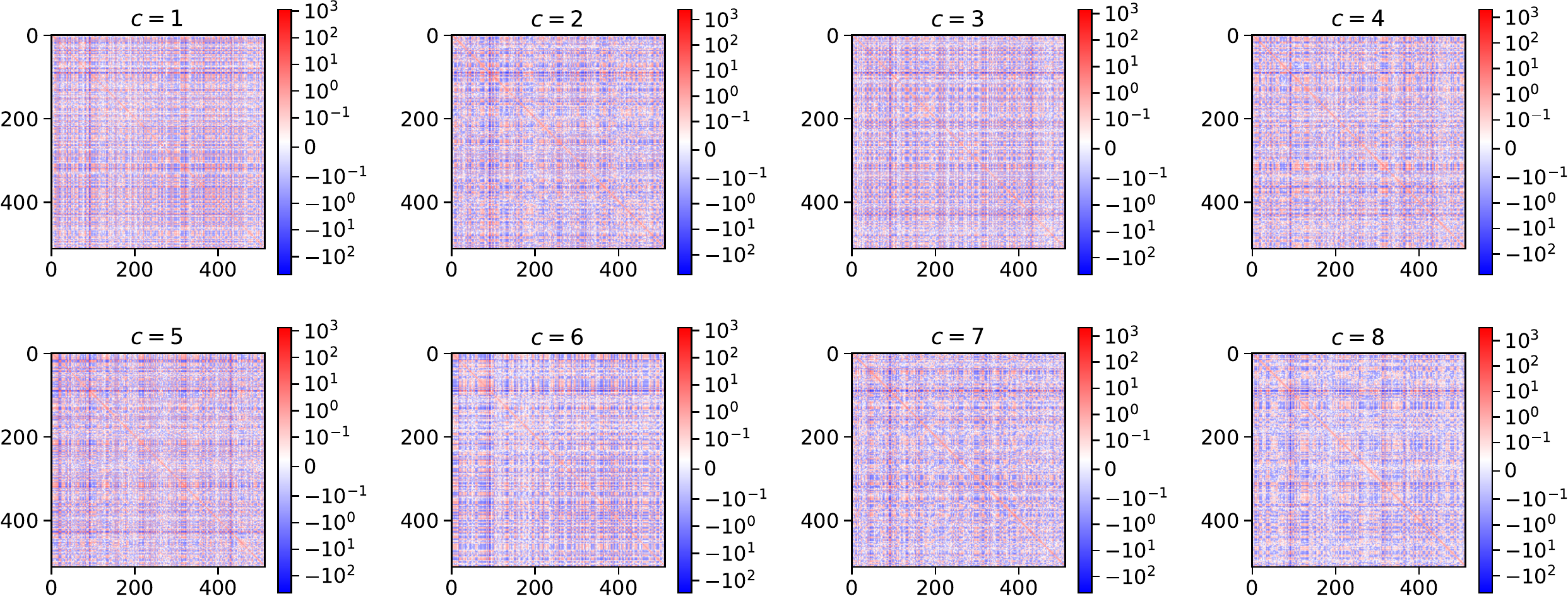}%
        \\[6pt]
        \includegraphics[width=\textwidth]{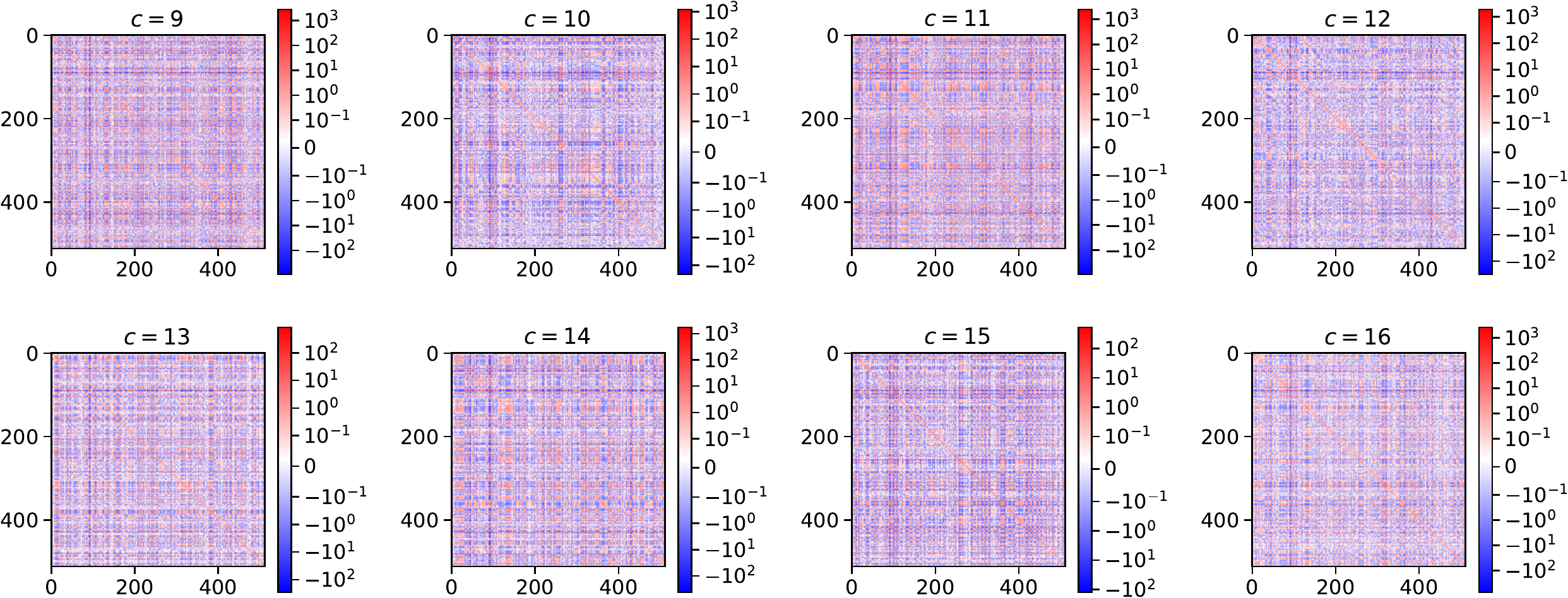}%
    \end{minipage}
    \hspace{0.5em}
    \begin{minipage}{0.3\textwidth}
        \includegraphics[width=\textwidth]{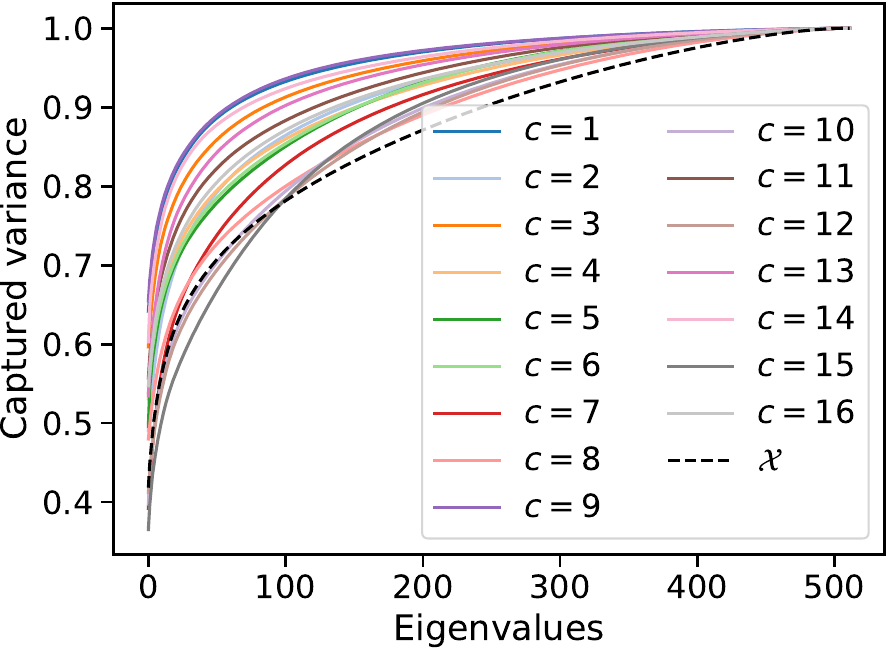}%
    \end{minipage}

    \caption{When clustering the LAION-1M-OOD dataset using spherical k-means into clusters $\{ \set{X}_c \}_{c=1}^{16}$, the correlations matrices $\sum_{\vect{x} \in \set{X}_c} \vect{x} \transpose{\vect{x}}$ of each cluster $\set{X}_c$ reveal a significant amount of structure in the data: they exhibit a checkerboard-like pattern, implying that their intrinsic dimensionality is lower than the full space $D$. This is corroborated when observing the captured variance (on the right), i.e., the cumulative sum of the eigenvalues, sorted in decreasing order. There, $\sim$150 target dimensions suffice to capture 80\% of the variance for every $c$. More importantly, the intrinsic dimensionality of each cluster is lower than that of $\set{X}$ as evidenced by the profile of its captured variance.}
    \label{fig:correlations}
\end{figure*}

We aim to assess the suitability of partitioning the database vectors using spherical k-means and applying linear dimensionality reduction on each cluster versus a direct application of the same linear algorithm. In \cref{fig:correlations}, we first compute 16 clusters. In their correlation matrices, we observe a clear checkerboard-like pattern, i.e., a sign of strong correlations between dimensions. When considering their profiles of captured variance, we observe that most of the variance in the vector cloud is contained in the first $\sim$150 principal directions. Note that variance profile of the entire set of vectors is upper bounded by that of the clusters (with one exception). This implies that partitioning and then applying a linear projection will capture more information from the database vectors than the application of a linear projection to the entire set.

\begin{figure*}
    \centering
    \includegraphics[width=\textwidth]{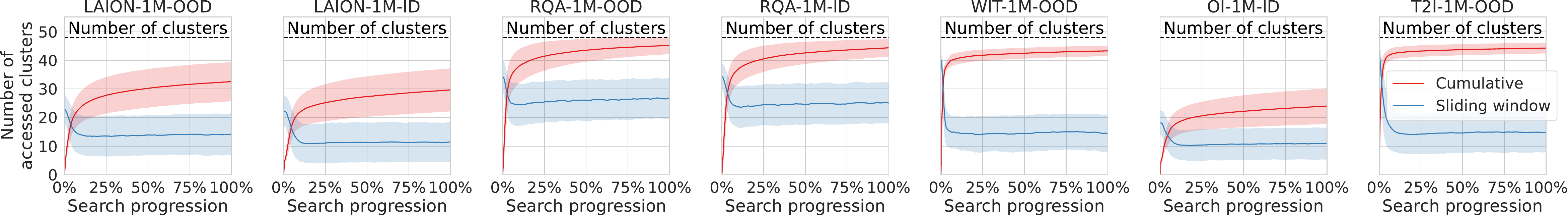}
    
    \caption{As the graph search progresses, the pattern of accessed cluster tags, see \cref{eq:cluster_tag}, favors the eager technique in \cref{algo:gleanvec_ip_eager} as it facilitates caching the queries $\{ \vect{q}_c \}_{c=1}^C$. The total number of visited tags (red curve) initially grows rapidly but this growth quickly slows down (notice the elbow at $\sim$10\%). In many cases, the search never visits all $C=48$ clusters. The number of visited tags in a sliding window over the search (blue curve) quickly reaches a steady-state and converges to a small fraction of $C$, which constitutes the elements of $\{ \vect{q}_c \}_{c=1}^C$ that we need to keep in cache to accelerate the search. For both curves, we show the mean across 1000 queries plus/minus one standard deviation.}
    \label{fig:search_history}
\end{figure*}

Next, we compare of the lazy and eager techniques for computing inner products (\cref{algo:gleanvec_ip_lazy,algo:gleanvec_ip_eager}) in \cref{fig:search_history}.
For this, we analyze the pattern of cluster tags, see \cref{eq:cluster_tag}, accessed throughout multiple graph searches. First, we observe that in many cases the search never visits all $C=48$ clusters (red curve). Additionally, the number of visited tags in a sliding window over the search (blue curve) quickly reaches a steady-state and converges to a small fraction of $C$, which constitutes the elements of $\{ \vect{q}_c \}_{c=1}^C$ that we need to keep in cache in \cref{algo:gleanvec_ip_eager}. All in all, this favors the eager algorithm, particularly when considering that its preprocessing step can be trivially parallelized.

Finally, we compare the search accuracy of GleanVec with that of LeanVec-Sphering (as the best linear technique). We evaluate their behavior under different target dimensionalities and varying the number of retrieved approximate neighbors. In all cases, GleanVec outperforms LeanVec-Sphering (the accuracy gap is wider for lower target dimensionalities). In most cases, we observe an advantage for using $C=48$ over $C=16$ clusters. This advantage is not too wide, though, and it comes with potentially more contention in cache placements when using the eager inner-product algorithm (\cref{algo:gleanvec_ip_eager}). Further experiments and benchmarks will help elucidate reasonable default values for the number $C$ of clusters.

\begin{figure*}
    \centering
    \begin{tblr}{
            rows={f},
            colsep=0pt,
            colspec={X[1,l,m] X[50,l]},
    }
        \begin{sideways}
            \scriptsize
            LAION-1M-OOD
        \end{sideways}
        & \includegraphics[width=0.98\textwidth]{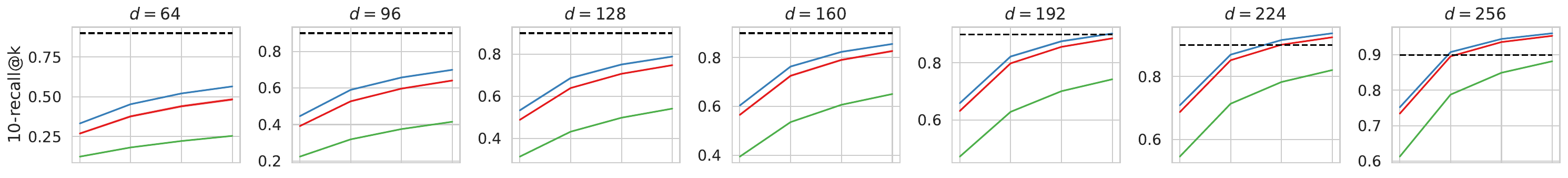} \\
        \begin{sideways}
            \scriptsize
            LAION-1M-ID
        \end{sideways}
        & \includegraphics[width=0.98\textwidth]{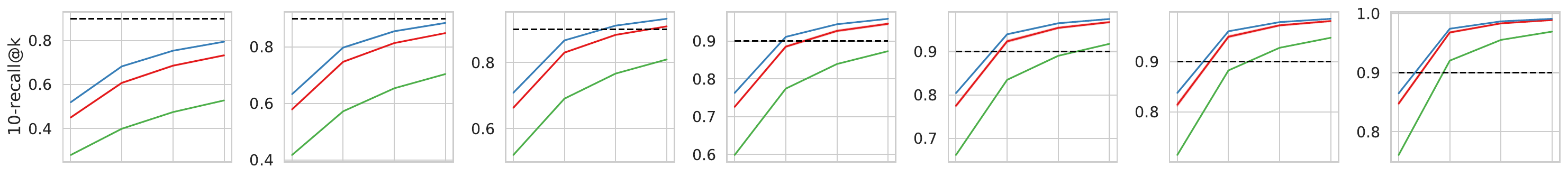} \\
        \begin{sideways}
            \scriptsize
            RQA-1M-OOD
        \end{sideways}
        & \includegraphics[width=0.98\textwidth]{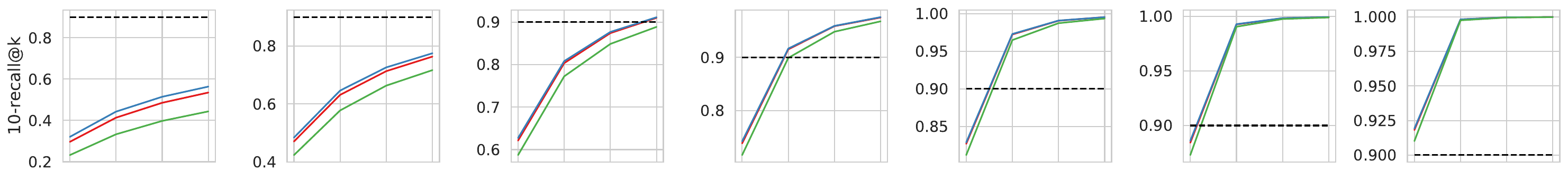} \\
        \begin{sideways}
            \scriptsize
            RQA-1M-ID
        \end{sideways}
        & \includegraphics[width=0.98\textwidth]{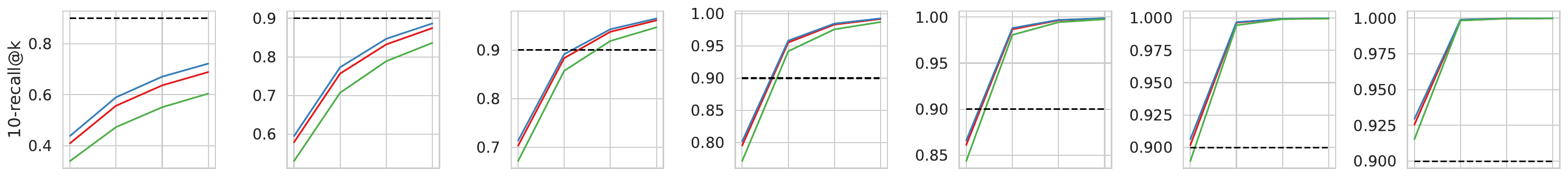} \\
        \begin{sideways}
            \scriptsize
            WIT-1M-OOD
        \end{sideways}
        & \includegraphics[width=0.98\textwidth]{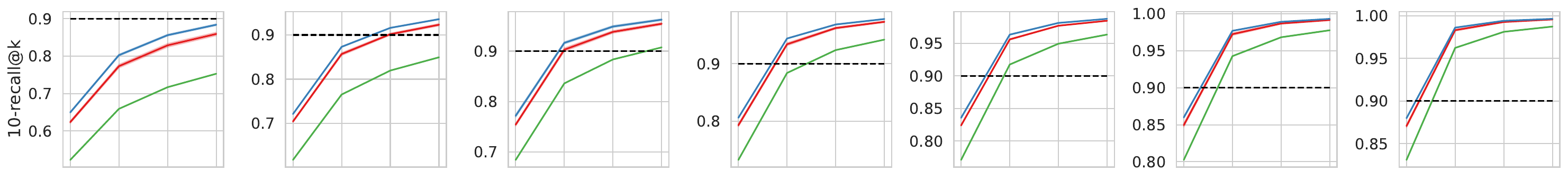} \\
        \begin{sideways}
            \scriptsize
            OI-1M-ID
        \end{sideways}
        & \includegraphics[width=0.98\textwidth]{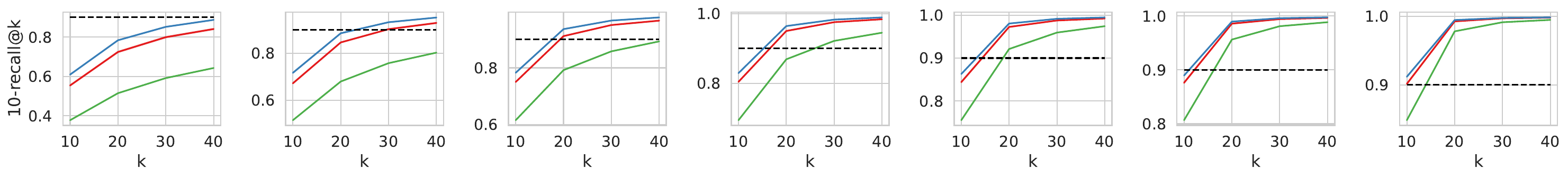} \\
        \begin{sideways}
            \scriptsize
            T2I-1M-OOD
        \end{sideways}
        & \includegraphics[width=0.98\textwidth]{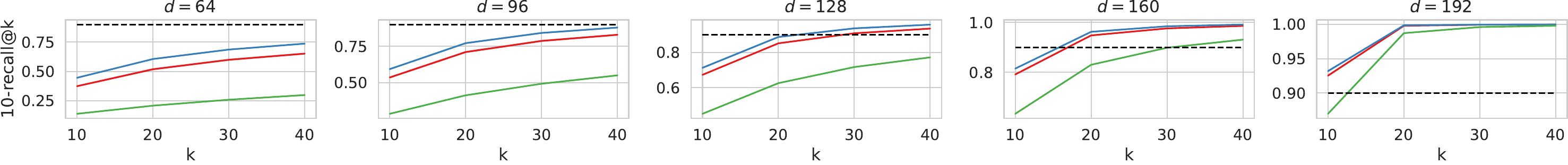} \\
    \end{tblr}
    
    \includegraphics[width=0.6\textwidth]{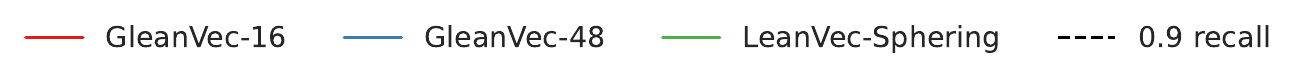}
    
    \caption{5 repetitions where made for GleanVec to account for the spherical k-means variation, we show plus/minus one standard deviation (barely visible). For T2I-1M-OOD, we explore fewer values of $d$ as $D=200$.}
    \label{fig:bruteForce}
\end{figure*}

\section{Related work}
\label[section]{sec:background}

Dimensionality reduction has a long standing use for approximate nearest neighbor search~\cite{deerwester1990indexing,ailon2009fast}. In this context, ID queries have been covered extensively in vector search~\cite{jegou_aggregating_2010,gong2012iterative,babenko2014inverted,wei_projected_2014} and more recently in retrieval-augmented language models~\cite{he2021efficient,izacard2020memory}, but OOD queries remain largely unexplored.
Dimensionality reduction is also deeply related to metric learning~\cite{bellet2013survey,kaya2019deep}. Any metric learned for the database vectors will be suitable for vector search when using ID queries. However, this does not hold with OOD queries. In a recent example, CCST~\cite{zhang_connecting_2022} uses transformers to reduce the dimensionality of deep learning embedding vectors and ultimately to accelerate vector search. However, the computational complexity of transformers precludes their usage for search and circumscribes their application to index construction.

Compressing vectors through alternative means has also received significant attention in the scientific literature.
Hashing~\cite{indyk1998approximate,jafari_survey_2021} and learning-to-hash~\cite{wang_survey_2018,luo2023survey} techniques often struggle to simultaneously achieve high accuracy and high speeds.

Product Quantization (PQ)~\cite{jegou2010product} and its relatives~\cite{ge2013optimized,babenko_additive_2014,zhang2014composite,andre_cache_2015,matsui_paper_2018,guo2020accelerating,wang_deltapq_2020,johnson_billion-scale_2021,andre2021quicker_adc,ko2021low} were introduced to handle large datasets in settings with limited memory capacity~\cite[e.g.,][]{jayaram2019diskann,jaiswal2022ood}.
With these techniques, (sub-segments of) each vector are assigned to prototypes and the similarity between (sub-segments of) the query and each prototype is precomputed to create a look-up table of partial similarities.
The complete similarity is computed as a sequence of indexed gather-and-accumulate operations on this table, which are generally quite slow~\cite{pase2019gather}. This is exacerbated with an increased dimensionality $D$: the lookup table does not fit in L1 cache, which slows down the gather operation even further.
Quicker ADC~\cite{andre2021quicker_adc} offers a clever fix by optimizing these table lookup operations using AVX shuffle-and-blend instructions to compute the similarity between a query and multiple database elements in parallel. This parallelism can only be achieved if the database elements are stored contiguously in a transposed fashion. This transposition, and Quicker ADC by extension, are ideally suited for inverted indices~\cite{johnson_billion-scale_2021} but are not compatible with the random memory access pattern in graph-based vector search.

\section{Conclusions}
\label[section]{sec:conclusions}

In this work, we presented two new linear and nonlinear methods for dimensionality reduction to accelerate high-dimensional vector search while maintaining accuracy in settings with in-distribution (ID) and out-of-distribution (OOD) queries. The linear LeanVec-Sphering outperforms other linear methods, trains faster, comes with no hyperparameters, and allows to set the target dimensionality more flexibly. The nonlinear Generalized LeanVec (GleanVec) uses a piecewise linear scheme to further improve the search accuracy while remaining computationally nimble. Initial experimental results show that LeanVec-Sphering and GleanVec push the state of the art for vector search.

In a revision of this manuscript, we will integrate LeanVec-Shpering and GleanVec into a state-of-the-art vector search library and assess the ensuing end-to-end performance benefits.

\printbibliography

\clearpage
\appendix

\section{Spherical k-means}
\label[appendix]{sec:spherical_kmeans}

Let $S^{D-1} = \left\{ \vect{x} \in \Real^D \ |\ \norm{\vect{x}}{2} = 1 \right\}$.
Let $\widetilde{\set{X}} = \left\{ \widetilde{\vect{x}}_i \in \Real^D \right\}_{i=1}^{n}$ be the normalized dataset, where each vector $\widetilde{\vect{x}}_i = \vect{x}_i / \norm{\vect{x}_i}{2}$.
Spherical k-means~\cite{dhillon2001concept} finds cluster centers $\{ \vect{\mu}_c \in S^{D-1} \}_{c=1}^C$ according to
\begin{equation}
    \max_{\{ \vect{\mu}_c \}_{c=1}^C} \sum_{i=1}^{n} \max_{c_i=1,\dots,C} \langle \widetilde{\vect{x}}_i \,,\, \vect{\mu}_{c_i} \rangle
    \quad\text{subject to}\quad
    \vect{\mu}_{c} \in S^{D-1}
    .
\end{equation}
As k-means, this problem is non-convex. A local minimum can be reached by iteratively applying the EM-like optimizations,
\begin{align}
    (\forall i)\ c_i &\gets \argmax_{c} \langle \widetilde{\vect{x}}_i \,,\, \vect{\mu}_{c} \rangle ,\\
    (\forall c)\ \vect{\mu}_c &\gets \left( \sum_{i=1}^{n} \indicator{c_i = c} \widetilde{\vect{x}}_i \right) / \norm{\sum_{i=1}^{n} \indicator{c_i = c} \widetilde{\vect{x}}_i}{2} .
\end{align}
To initialize the cluster centers $\{ \vect{\mu}_c \}_{c=1}^C$, we use k-means++ on $\widetilde{\set{X}}$. Alternatively, depending on the dataset distribution, we could use the Pyramid Vector Quantizer~\cite{fischer1986pyramid,duda2017improving}

In practice, since $C$ is a small number ($C<100$), we compute the cluster centers $\{ \vect{\mu}_c \}_{c=1}^C$ from a uniform random sample of $\widetilde{\set{X}}$ containing $10^5$ points. Alternatively, we could use a coreset~\cite[e.g.,][]{bachem2018one} to increase the quality of the clustering. We have not observed any catastrophic effects from uniform sampling and we thus select this simpler option.

\subsection{Adapting other similarities}
\label[appendix]{sec:adapting_similarities}

For the datasets whose ground truth was computed using Euclidean distance (see \cref{tab:datasets_ground_truth_metric}), we adapt the vectors to the inner product setting covered in this work. For the datasets that have (nearly) normalized vectors, minimizing Euclidean distances is (almost) equivalent to maximizing inner products. For those with unnormalized vectors (e.g., GIST-1m-ID), we use the transformation
\begin{equation}
    \hat{\vect{x}} =
    \begin{bmatrix}
        \vect{x} \\
        -\tfrac{1}{2} \norm{\vect{x}}{2}^2
    \end{bmatrix},
    \quad
    \hat{\vect{q}} =
    \begin{bmatrix}
        \vect{q} \\
        1
    \end{bmatrix} .
\end{equation}
As a consequence, the transformed query and database vectors have one additional dimension (totalling $\hat{D}=961$ in GIST-1M-ID).
Then,
\begin{align}
    \argmax_{i} \left\langle \hat{\vect{q}}, \hat{\vect{x}}_i \right\rangle
    &=
    \argmax_{i}
    \left\langle \vect{q}, \vect{x}_i \right\rangle
    - \tfrac{1}{2} \norm{\vect{x}_i}{2}^2
    \\
    &=
    \argmax_{i}
    \left\langle \vect{q}, \vect{x}_i \right\rangle
    - \tfrac{1}{2} \norm{\vect{x}_i}{2}^2
    - \tfrac{1}{2} \norm{\vect{q}}{2}^2
    \\
    &=
    \argmax_{i}
    -\tfrac{1}{2} \norm{ \vect{q} - \vect{x}_i}{2}^2
    \\
    &=
    \argmin_{i}
    \norm{ \vect{q} - \vect{x}_i}{2}^2 ,
\end{align}
allowing to use inner product on the transformed vectors while obtaining results that are compatible with the ground truth Euclidean dissimilarity.

\section{Datasets}
\label[appendix]{sec:datasets}

We evaluate the effectiveness of the proposed methods on a wide range of in-distribution (ID) and out-of-distribution (OOD) datasets as shown in \cref{table:datasets}.

\noindent\textbf{ID datasets:}
\begin{itemize}
    \item We use GIST-1M and DEEP-1M, two standard ID datasets~\cite{zhang_connecting_2022}.\footnote{\url{https://www.cse.cuhk.edu.hk/systems/hash/gqr/datasets.html}} We utilize the learn sets provided in these datasets to construct test and validation query sets, with the first 10K entries as test and the next 10k as validation.
    \item We use the ID datasets OI-1M and OI-13M~\cite{aguerrebere2024high}, with 1 million and 13 million database vectors, generated from a subset Google's Open Images~\cite{OpenImages} using the OpenAI CLIP-ViT-B32 model~\cite{radford2021learning}.
    \item We use an instance with 1M vectors of a dataset~\cite{tepper2023leanvec} with ID queries stemming from a question-answering application. Here, the vectors are text encoded using the RocketQA dense passage retriever model~\cite{qu-etal-2021-rocketqa}. Both the query and the database vectors are taken from a common pool of text answers, encoded with the RocketQA dense passage retriever model~\cite{qu-etal-2021-rocketqa}.\footnote{\label{footnote:rocketQA}\url{https://github.com/PaddlePaddle/RocketQA}}
\end{itemize}

\noindent\textbf{OOD datasets:}
\begin{itemize}
    \item We use three cross-modal text-to-image datasets, namely T2I-1M~\cite{text2image}, T2I-10M~\cite{text2image}, and LAION-1M~\cite{Schuhmann2021}, where the query and database vectors are text and image embeddings, respectively. In all three datasets, the queries are divided into learning a test sets.
    \item We use WIT-1M, a dataset~\cite{tepper2023leanvec} with OOD queries stemming from a text-to-image application.
    The query set is built using text descriptions and generating the corresponding embeddings using CLIP-ViT-B32-multilingual-v1~\cite{Reimers2020}. The database vectors are built from the images, encoded with the multimodal OpenAI CLIP-ViT-B32 model~\cite{radford2021learning}.
    \item We use two instances of a dataset~\cite{tepper2023leanvec} with OOD queries stemming from a question-answering application, RQA-1M and RQA-10M, respectively with 1M and 10M vectors. Here, the vectors are text encoded using the RocketQA dense passage retriever model~\cite{qu-etal-2021-rocketqa}.\footnote{See footnote \ref{footnote:rocketQA}}
    The OOD nature of the queries emerges as dense passage retrievers use different neural networks to encode the questions (i.e, queries) and the answers (i.e., database vectors).
\end{itemize}

\begin{table}
    \caption{The datasets in \cref{table:datasets} and \cref{sec:datasets} along with the similarities and vector normalization used to compute their ground truth. Note that the Euclidean distance is a dissimilarity. The adaption method is described in \cref{sec:adapting_similarities}.}
    \label[table]{tab:datasets_ground_truth_metric}

    \small
    \centering
    \begin{tblr}{lccc}
        \hline
        Dataset & Ground truth (dis)similarity & Normalized vectors & Adapted to IP \\ 
        \hline
        GIST-1M-ID & Euclidean & No & Yes \\
        DEEP-1M-ID & Euclidean & Approximately (3\% variation) & No \\
        OI-1M-ID & Euclidean / cosine & Yes & No \\
        OI-13M-ID & Euclidean / cosine & Yes & No \\
        RQA-1M-ID & Inner product & No & No \\
        T2I-1M-OOD & Inner product & No & No \\
        T2I-10M-OOD & Inner product & No & No \\
        LAION-1M-OOD & Inner product & Yes & No \\
        WIT-1M-OOD & Inner product & No & No \\
        RQA-1M-OOD & Inner product & No & No \\
        RQA-10M-OOD & Inner product & No & No \\
        \hline
    \end{tblr}
\end{table}

\end{document}